\documentclass[slac_one]{revtex4}
\usepackage{graphicx}
\usepackage{fancyhdr}
\begin{document}

\title{Reconstruction and Identification of Hadronic Decays of Taus using the CMS Detector} 

%

\author{E.K. Friis (for the CMS collaboration)}
\affiliation{UC Davis, Davis, CA 95616, USA}
%

\begin{abstract}
New physics beyond the Standard Model could well preferentially
show up at the LHC in final states with taus. The development of
efficient and accurate reconstruction and identification of taus is
therefore an important item in the CMS physics programme. The potentially
superior performance of a particle-flow approach can help to achieve this
goal with the CMS detector. Preliminary strategies are presented in
this summary for the hadronic decays of the taus.
\end{abstract}

\maketitle

\thispagestyle{fancy}




\section{INTRODUCTION}
Because the tau is the heaviest of the three leptons, specific final states involving taus are expected to show up 
in the Standard Model (SM), and would appear abundantly in many processes arising from new physics beyond the SM. 
Because taus decay to hadronic final states in about 64\% of the cases, the reconstruction and identification of these hadronic decays 
is an essential ingredient of the Compact Muon Solenoid (CMS) physics programme. The CMS tau identification uses objects from the particle-flow reconstruction, 
which combines information from all CMS subdetectors to produce a global event description at the level of individually reconstructed particles.
The individual particle list includes muons, 
electrons (with individual bremsstrahlung photons), photons (unconverted or
converted), charged hadrons (without or with a nuclear interaction in the tracker), and stable and unstable 
neutral hadrons. The details of the particle-flow algorithm and its full implementation in CMS will be described in a 
forthcoming paper~\cite{pflow}.
\section{TAU RECONSTRUCTION}
Unlike in the case of electrons and muons, hadronic taus cannot use a generic tag-and-probe technique to measure efficiency from data. 
One option is to apply a basic selection based on loose isolation to reduce the huge QCD background keeping a large efficiency for all tau decay modes.
Any possible loss of efficiency in the basic selection due to the presence of 
pileup or underlying particles can be determined from the data via ${\rm Z} \to {\rm e}{\rm e}(\mu\mu)$ events, providing a relatively pure sample
from which the performance of higher-level algorithms can be measured.
\subsection{Base selection algorithm}
First, a transverse-momen\-tum threshold is applied to each tau candidate jet. 
Next, the highest $p_{\rm T}$ (``leading'') track within $\Delta R < 0.1$ of the jet axis 
is required to have $p_{\rm T}$ above 5\,GeV/$c$.
A narrow ``signal cone'', expected to contain all tau decay products, 
and an ``isolation annulus'', expected to contain little activity if the
tau is indeed isolated, 
are defined about the leading track.
To enforce the tau isolation, no
reconstructed charged hadrons with $p_{\rm T}$ above 1 GeV/$c$ and no photons with $p_{\rm T}$ larger than 1.5 GeV/$c$ 
are allowed in the isolation annulus.

\begin{figure}[pht]
\centering
      \includegraphics{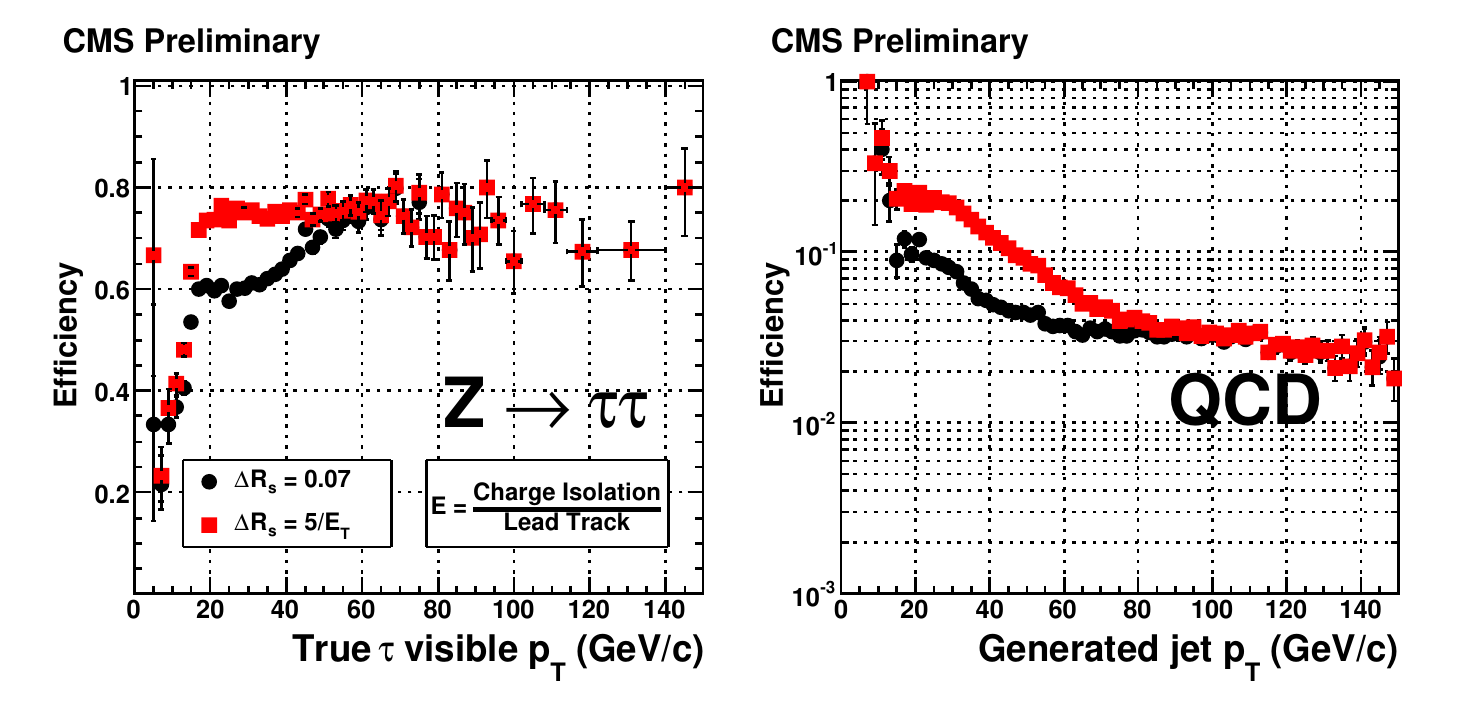}
   \caption{Comparison of the marginal efficiencies for the charged-hadron isolation requirement efficiency as a function of the true 
   visible $p_{\rm T}$ for taus (left) and QCD jets (right) with the shrinking ($\Delta R_{\rm sig} < 5/E_{\rm T}$) and 
   fixed ($\Delta R_{\rm sig} = 0.07$) signal-cone definitions.}
   \label{fig:isolationFixedvs5ETPt}
\end{figure}

Up to now, many CMS analyses~\cite{ptdr2} have used fixed-sized signal
and isolation cones of typical sizes $\Delta R$ = 0.07
and 0.45, respectively.
An
alternative 
is to utilize the fact that high energy taus are Lorentz boosted and
hence become more collimated at higher energy:  the ``signal'' cone size is
defined to shrink inversely proportional to the jet transverse energy,
$5/E_T$, within the limits of 0.07 and 0.15.
A comparison of the performance of the (historical) $\Delta R_{\rm sig}
= 0.07$ and the (new) $\Delta R_{\rm sig} = 5/E_{\rm T}$ signal cones
in terms of the marginal efficiency of the charged hadron isolation
requirements is shown in
Fig.~\ref{fig:isolationFixedvs5ETPt} for taus and QCD jets as a function
of $p_{\rm T}$.  
An increase of approximately 20\% is observed in the signal
efficiency
for the low-$p_{\rm T}$ region, with an approximate doubling of the
background rate. The improved efficiency of the shrinking-cone
algorithm is due to a better acceptance of three-prong taus in the
low-to-intermediate $p_{\rm T}$ range.
The recovery of the three-prong decays is essential to make the base selection independent of the decay mode, and allow the use of
higher level techniques to further suppress the background.  

\begin{figure}[pht]
\centering
      \includegraphics{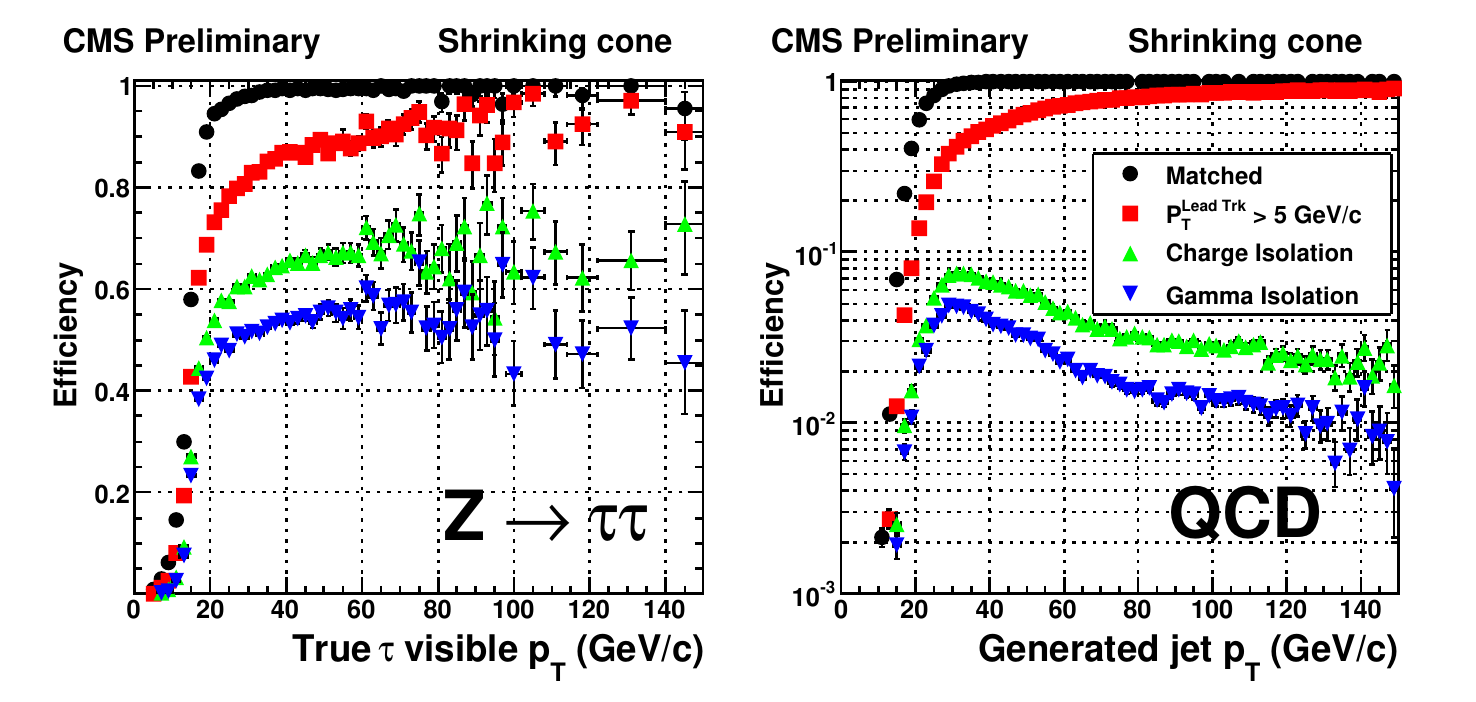}
   \caption{Global efficiencies of the successive base selection cuts for taus (left) and QCD jets (right) as a function of the true visible $p_{\rm T}$
   for the $5/E_{\rm T}$ shrinking-cone algorithm.}
   \label{fig:valStylePtHE}
\end{figure}

Photon isolation is another powerful discriminator against QCD jet
backgrounds.  
Since the substantial amount of tracker material leads to
high rate of
photon conversions, low-energy electrons from photon conversions may appear as
photons in the isolation annulus.
Work is ongoing in the particle flow group to reconstruct the resulting secondary tracks, and will 
allow a re-optimization of the photon signal cone definition.
The efficiencies of each base selection step (jet-finding, lead track finding, charge isolation and gamma isolation) for the new ``shrinking-cone'' algorithm
is given in Figure~\ref{fig:valStylePtHE}.  A full comparison to the historical ``fixed-cone'' can be found in~\cite{tauPas}.
\subsection{Higher level identification\label{highLevel}}
Following the base tau reconstruction, 
a high level 
identification stage is designed to achieve higher purity samples suitable for individual physics analyses.  Electron and muon rejection algorithms have been implemented,
and higher-level QCD discriminants utilizing the extra information provided by the particle-flow reconstruction are being actively developed.

\begin{figure}[pht]
      \includegraphics[width=0.5\textwidth]{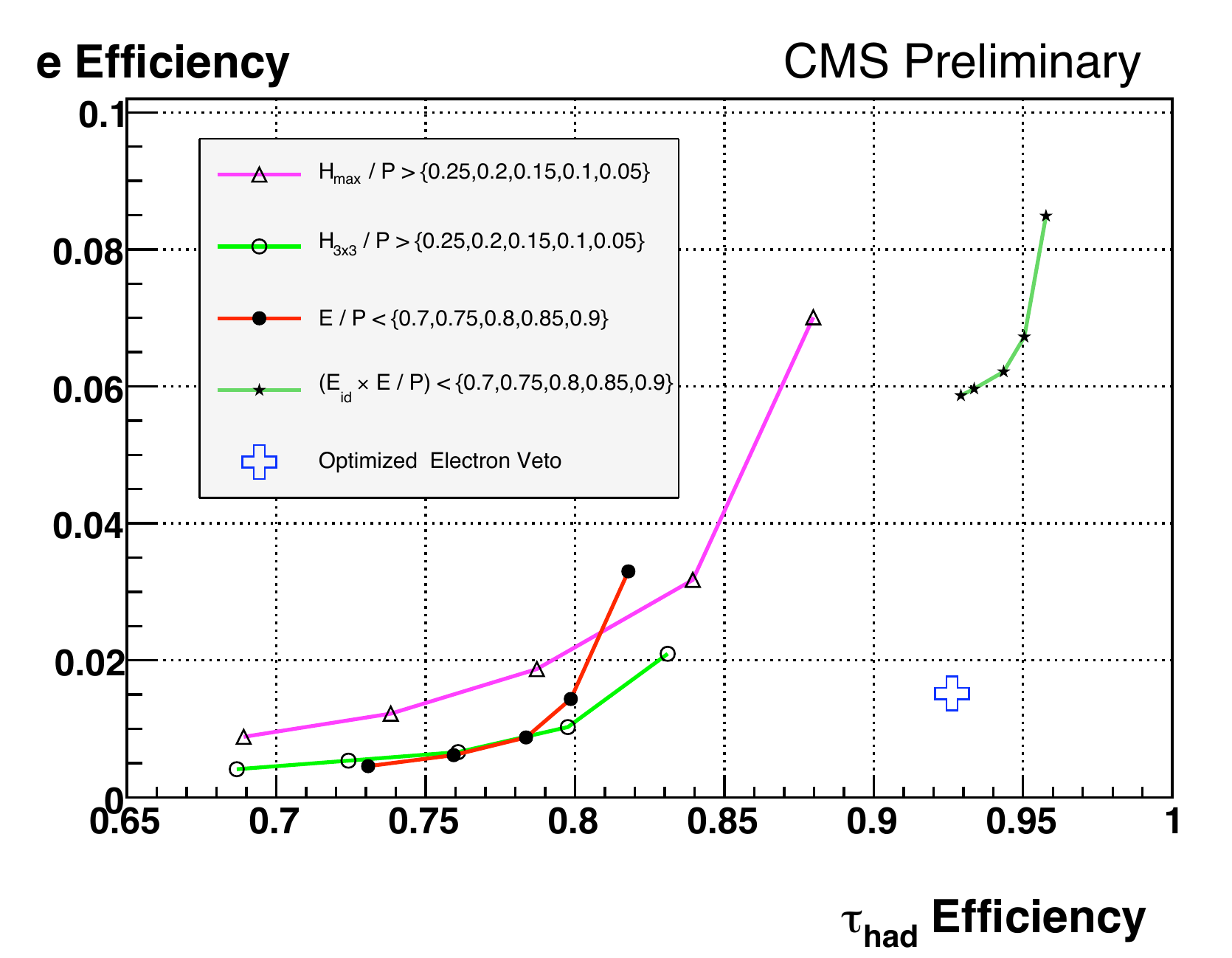}
   \caption{Electron vs. tau selection efficiency for several typical electron rejection approaches compared 
   to the Optimized Electron Veto. 
   The quantities used are fully defined in~\cite{tauPas}.}
   \label{fig:EffvsRej}
\end{figure}

Isolated electrons produced in the electroweak processes, e.g. ${\rm Z} \to {\rm ee}$, 
become an important source of misidentified taus in many physics analyses. 
A particle-flow electron pre-identification algorithm has been developed, described in \cite{pflow} and
achieves 90-95\% efficiency
with about 5\% pion efficiency.  
To extend the electron rejection beyond 95\%, two additional variables are formed, which are described fully in \cite{tauPas}. 
The first variable, E/P, is the energy of ECAL clusters expected to contain electron 
bremsstrahlung photons divided by the momentum of the leading track.
The second variable, H$_{3\times3}$/P, is the energy of HCAL clusters expected to contain the charged pion shower divided by the momentum of the leading track.
Optimizing cuts on these variables separately for taus which are pre-identified as electrons as those that are not 
leads to an efficiency of 92.5\% for true taus and 1.5\% for true electrons. 
A summary of all results is shown in Fig.~\ref{fig:EffvsRej}, with the optimized cuts described above labeled as ``Optimized Electron Veto''.
\section{CONCLUSION}
This summary describes tau reconstruction and identification using particle flow with the CMS detector. 
There are three major components: a general particle flow reconstruction, a common tau reconstruction using reconstructed particles, 
and a higher level identification.  
While further significant improvements 
are still being pursued, existing methods provide a strong rejection of QCD, electron and muon backgrounds, while
preserving high efficiency for selecting hadronic taus.
\end{document}